\documentclass[pre,floatfix,amsmath,twocolumn]{revtex4}

\usepackage{graphicx}
\usepackage{amsmath}
\usepackage{amssymb}
\usepackage{calrsfs}

\newcommand{\rombarrier}{{\operatorname{barrier}}}
\newcommand{\romB}{{\operatorname{B}}}
\newcommand{\romc}{{\operatorname{c}}}
\newcommand{\romd}{{\operatorname{d}}}
\newcommand{\romfree}{{\operatorname{free}}}

\newcommand{\CALO}{{\mathcal O}}

\newcommand{\VECr}{{\boldsymbol{r}}}

\newcommand{\eg}{{\emph{e.}$\,$\emph{g.}}}
\newcommand{\ie}{{\emph{i.}$\,$\emph{e.}}}

\newcommand{\un}[1]{\,\text{#1}}

\begin{document}


\title{Wrapping of a spherical colloid by a fluid membrane}
\author{Markus Deserno$^1$}
\author{Thomas Bickel$^{1,2}$}
\affiliation{$^1$Department of Chemistry and Biochemistry, UCLA, %
             Box 951569, Los Angeles, CA 90095, USA}
\affiliation{$^2$Department of Physics and Astronomy, UCLA, %
             Box 951547, Los Angeles, CA 90095, USA}

\date{December 16, 2002}


\begin{abstract}
  We theoretically study the elastic deformation of a fluid membrane
  induced by an adhering spherical colloidal particle within the
  framework of a Helfrich energy.  Based on a full optimization of the
  membrane shape we find a continuous binding and a discontinuous
  envelopment transition, the latter displaying a potentially
  substantial energy barrier.  A small gradient approximation permits
  membrane shape and complex energy to be calculated
  ana\-ly\-ti\-cal\-ly.  While this only leads to a good
  representation of the complex geometry for very small degrees of
  wrapping, it still gives the correct phase boundaries in the regime
  of low tension.
\end{abstract}


\maketitle


\section{Introduction and Motivation}

Fluid lipid bilayers are one of the key structural elements of all
living cells. On the one hand they partition and thereby organize the
complex hierarchy of intracellular biochemical environments.  On the
other hand they provide controlled transport between neighboring
compartments as well as the extracellular space.  These transport
mechanisms span quite a large range of particle sizes, all the way
from sub-nanometer ions (which cross the membrane via protein
channels) up to micron sized objects (like bacteria), which are
engulfed in large-scale membrane deformations occurring during
phagocytosis \cite{Lodish}.  Generally, such events require metabolic
energy and are meticulously controlled by the cell.  However, there
are also cases where they happen passively as a consequence of generic
physical interactions, for instance a sufficiently strong adhesion
between the particle about to be transported and the membrane.  An
important and well studied example is provided by the route along
which many animal viruses leave their host cells, namely, via the
wrapping and subsequent pinch-off of a pre-assembled viral
nucleoprotein capsid at the plasma membrane \cite{GaHe98}.  Other
examples, in which membrane deformations following binding to a small
object are crucial, include certain gene transfection systems, in
which DNA is complexed by positively charged polymers and the
resulting condensed globule becomes internalized by the cell in an
adhesion-driven invagination process \cite{BoLe95}; or a host of
modern biophysical experimental techniques, in which for instance
microbeads \cite{CaYe01} or AFM tips \cite{HeOb00} are in contact with
a membrane and become partially or fully wrapped.

These biological examples are complemented by more physically oriented
experiments on the adsorption of micrometer sized beads onto model
lipid bilayers. Studies have for instance focused on the wrapping of a
latex bead by a vesicle \cite{DiAn97}, or on the interaction between
several beads via membrane mediated forces \cite{KoRa99}.  For these
non-flat substrates, it is unfortunately difficult to extract detailed
information about the membrane shape close to contact, especially for
small colloids for which the bending contribution in the wrapping
balance becomes more important \cite{DeGe02}.  It is therefore
desirable to have a better theoretical understanding of how physical
parameters like bending stiffness, lateral tension, or adhesion
strength control the shape of the complex and under which
circumstances complete wrapping ensues.  This question has been
partially addressed in a recent work that considers the adhesion of a
cylindrical rod to a membrane and estimates the force required for
unbinding \cite{Bou02}.  However, the description is limited to the
regime of small wrapping, since it relies on a small gradient
expansion of the elastic energy.  In the present paper we determine
the complete structural wrapping behavior of a spherical colloid by
numerically solving the full nonlinear differential equations which
describe the shape of an elastic fluid membrane under a prescribed
lateral tension adhering to a spherical particle with some given
strength.  These exact results are then compared to analytical
calculations employing a small gradient approximation.  We find that
in the latter case the shape of the complex is only predicted
correctly for very small degrees of wrapping.  However, the phase
boundary toward envelopment is still accurately represented over a
somewhat larger range of penetrations, essentially in the regime of
low tension.


\section{General energy considerations}

\begin{figure}
\includegraphics[scale=0.7]{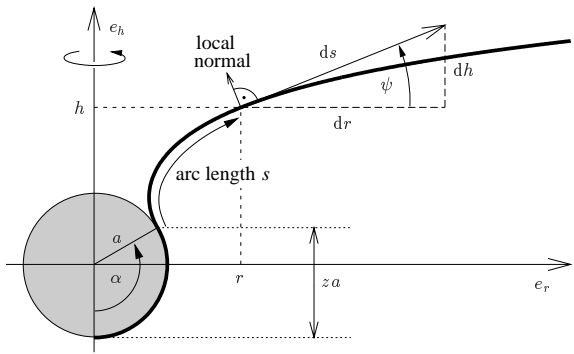}
\caption{Geometry of the wrapping complex and membrane parameterization.
  A membrane adheres partially and cylindrically symmetrically to a
  spherical colloid of radius $a$ with a degree of wrapping given by
  $z=1-\cos\alpha$.  Due to the possibility of ``overhangs'' it is
  advantageous to parameterize the membrane by specifying the angle
  $\psi$ as a function of arc length $s$.  The more direct choice of
  measuring the height $h$ as a function of radial distance $r$
  remains however useful for the small gradient approximation.}
\label{fig:definitions}
\end{figure}

We consider a spherical colloid of radius $a$ adsorbed on a deformable
surface, as depicted in Fig.~\ref{fig:definitions}. Following
Helfrich~\cite{Hel73}, the elastic energy associated with some
membrane deformation is, per unit area,
\begin{equation}
  e_{\text{H}} = \sigma +\frac{\kappa}{2}(c_1+c_2)^2 \ ,
\end{equation}
with $c_1$ and $c_2$ the local principal curvatures, $\kappa$ the
bending modulus, and $\sigma$ the lateral tension. From the two elastic
constants we can construct a length, $\lambda$, according to
\begin{equation}
  \lambda = \; \sqrt{\frac{\kappa}{\sigma}} \ .
\end{equation}
For instance, with a tension of $\sigma \simeq 0.02\un{dyn/cm}$ (a
value commonly found for cell membranes \cite{MoHo01}) and a typical
bending modulus of $\kappa \simeq 20\,k_\romB T$ (where $k_\romB T$ is
the thermal energy) we obtain $\lambda \simeq 64\un{nm}$ (which,
incidentally, is about the same size as many viral nucleocapsids
\cite{BaOl99}).  Membrane deformations on a length scale smaller than
$\lambda$ are mainly controlled by bending energy, while tension is
predominant on larger scales.  In this study, we will restrict
ourselves to the bending-dominated low tension regime, $a \lesssim
\lambda$.  We will see that even then a comparatively low tension can
play an important role.

A natural parameter specifying the coverage of the colloid by the
membrane is the degree of wrapping $z=1-\cos \alpha$. It ranges from
$z=0$, when the colloid just touches the surface, up to $z=2$, in the
fully enveloped state. The equilibrium value of $z$ results from the
adhesion of the colloid, driven by a contact energy per unit area,
$w$, being balanced by the requirement to bend the membrane as well as
the work of pulling excess membrane toward the wrapping site against
the prescribed lateral tension $\sigma$.

Energies pertaining to the \emph{adhering} part can be calculated
easily: The adhesion energy equals $-2\pi a^2 z w$, while bending and
tension contributions are found to be $4\pi z\kappa$ and $\pi
a^2z^2\sigma$, respectively. The bending and tension energies of the
\emph{free} part of the membrane are more difficult to evaluate, since
they require the determination of the equilibrium membrane profile, a
task that will be pursued below.  Before solving the full variational
problem, it is worth noting that the situation is greatly simplified
for a membrane with no lateral tension. In this special case the
equilibrium profile is quickly seen to be a \emph{catenoid}, a minimal
surface with zero mean curvature.  Hence, the only energy
contributions stem from the wrapped part of the membrane.  We then
find that, for $\sigma=0$, colloids do not adhere at low adhesion
energy $w<w_\romc=2\kappa/a^2$, whereas full wrapping occurs above
$w_\romc$, with no energy barrier to be overcome.


\section{Nonlinear shape equations}

As a first approach, it would be tempting to approximate the energy of
the free section of the membrane by a phenomenological line energy.
However, neither the relation between the line tension constant and
the membrane properties $\kappa$ and $\sigma$ would be known, nor is
the implied dependency on the degree of wrapping correct.  In order to
draw indubitable conclusions, it is thus advisable to determine the
exact membrane profile.

The energy $E_\romfree$ of the free membrane is the surface integral
over the local bending and tension contributions and is thus a
\emph{functional} of the shape.  By solving the corresponding
Euler-Lagrange equations one obtains the ``ground state'' profile and
thereby its energy.  Much work along these lines has for instance led
to a detailed understanding of vesicle conformations \cite{Sei97}.  In
the present context, and using the parameterization indicated in
Fig.~\ref{fig:definitions}, the shape equations can be written as
\begin{subequations}
\label{eq:Hamilton}
\begin{eqnarray}
  \dot\psi
  & = &
  \frac{p_\psi}{2r}-\frac{\sin\psi}{r} \ ,
  \\
  \dot r
  & = &
  \cos\psi \ ,
  \\
  \dot h
  & = &
  \sin\psi \ ,
  \\
  \dot p_\psi
  & = &
  \Big(\frac{p_\psi}{r}-p_h\Big)\cos\psi +
  \Big(\frac{2r}{\lambda^2} + p_r\Big)\sin\psi \ ,
  \\
  \dot p_r
  & = &
  \frac{p_\psi}{r} \Big(\frac{p_\psi}{4r}-\frac{\sin\psi}{r}\Big) +
  \frac{2}{\lambda^2}(1-\cos\psi) \ ,
  \\
  \dot p_h
  & = &
  0 \ .
\end{eqnarray}
\end{subequations}

\begin{figure}
\includegraphics[scale=0.87]{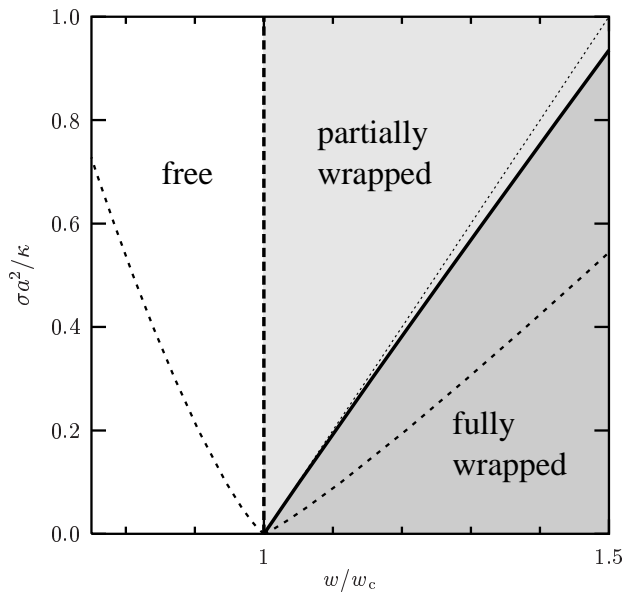}
\caption{Structural wrapping phase diagram in the plane of adhesion
  constant $w$ and lateral tension $\sigma$.  The bold solid line
  indicates the discontinuous transition between partially and fully
  wrapped, and the short dashed lines are the spinodals belonging to
  it.  The fine dotted line $\sigma=w-w_\romc$ indicates where the
  fully enveloped state has zero energy.}\label{fig:pd}
\end{figure}

The dot indicates the derivative with respect to the arc length $s$
along the membrane profile, and the $p$'s are the ``momenta''
canonically conjugate to the coordinates $\psi$, $r$, and $h$.  These
equations are integrated numerically for an asymptotically flat
membrane that touches the colloid tangentially at a specific point of
detachment. The determination of the appropriate boundary conditions
far from the colloid raises a few subtle technical questions, and we
refer the reader to Refs.~\cite{JuSe94} and \cite{Des} for further
details.  From the numerical solution we finally obtain $E_\romfree$
-- and thus the total energy -- as a function of the degree of
wrapping $z$.

\begin{figure}
\includegraphics[scale=1]{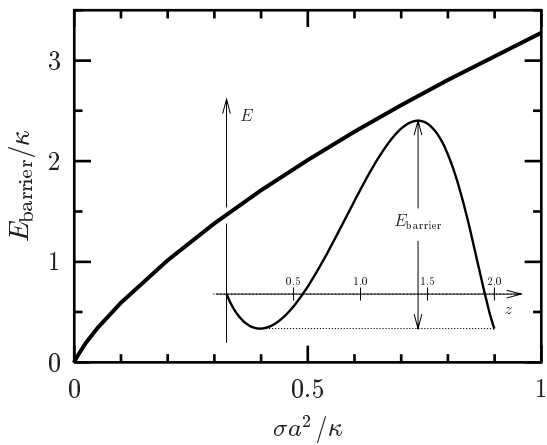}
\caption{Height of the energy barrier $E_\rombarrier$ as a function
  of reduced tension.  The inset illustrates the shape of the function
  $E(z)$ at the value of $w$ where the transition from partial to full
  wrapping occurs and illustrates the definition of the concomitant
  energy barrier.}\label{fig:barrier}
\end{figure}

The structural phase diagram depicted in Fig.~\ref{fig:pd} follows
directly.  We find that the transition from the free to the partially
wrapped state is continuous and occurs at the same value
$w_\romc=2\kappa/a^2$ for any value of the tension.  In contrast to
that, an energy barrier separates the partially wrapped from the the
fully enveloped state, rendering this transition discontinuous.  The
transition lines merge at a triple point $(w_\romc,0)$, as expected.
Our approach also allows us to determine the \emph{height} of the
energy barrier (see Fig.~\ref{fig:barrier}), which surprisingly
originates predominantly from tension.  For $\sigma=0.02\un{dyn/cm}$
(typical cellular tension \cite{MoHo01}), $a=30\,$nm (capsid radius of
Semliki Forest Virus, an often studied example) and
$\kappa=20\,k_\romB T$, the barrier has the substantial value of
$E_\rombarrier\approx 22\,k_\romB T$, showing that it cannot be
overcome by thermal fluctuations.  This point is also illustrated in
Fig.~\ref{fig:pd} by showing the two spinodal lines at which the
barrier vanishes.  Incidentally, starting with a fully enveloped state
and reducing the adhesion $w$, the unwrapping-spinodal is crossed only
once one is already in the unbound regime, \ie, the hysteresis would
be so pronounced that one ``skips'' the partially wrapped region upon
unbinding.

The energy of the free section of the membrane vanishes in the limit
of full wrapping, $z\rightarrow 2$, which is reminiscent of the case
of an ideal neck connecting two vesicles \cite{FoMi94}.  The reason is
essentially that as the neck contracts, the necessary membrane
back-bending must occur on a length scale much smaller than $\lambda$
and is thus entirely dominated by the bending energy.  But then the
membrane will just assume the shape of a catenoid which, even though
highly deformed, has zero mean curvature and thus does not cost any
bending energy.


\section{Small gradient expansion, rigorous results}

The exact shape equations~(\ref{eq:Hamilton}) are nonlinear and can
only be solved numerically.  In order to get analytical information
about the triple point at $(w_\romc,0)$, we use a single-valued
(Monge-) parameterization of the surface profile $h(\VECr)$, where
$\VECr=(x,y)$ spans the flat reference plane at $h=0$.  This
representation does not allow for overhangs and therefore can only
describe the first stages of the wrapping process.  Assuming moreover
that membrane deflections out of the horizontal remain small,
\ie, $\vert \nabla h \vert \ll 1$, the energy of the free part of
the membrane is
\begin{equation}
  \label{functional} E_\romfree = \int \romd^2 \VECr \left\{
  \frac{\kappa}{2} \left(\nabla^2 h\right)^2+\frac{\sigma}{2}
  \left(\nabla h\right)^2 \right\} \ .
\end{equation}
This quadratic expansion is valid only for small deformations but it
has the advantage to render analytical calculations tractable. The
equilibrium profile arises from the stationarity condition $\delta
E_\romfree/\delta h=0$ and satisfies the
\emph{linear} Euler-Lagrange equation
\begin{equation}
\label{euler}
\nabla^2 \left(\nabla^2 -\lambda^{-2}\right)h=0  \ .
\end{equation}
The general solution of eq.~(\ref{euler}) is $h(r) = h_1 +
h_2\ln(r/\lambda) + h_3K_0(r/\lambda) + h_4 I_0(r/\lambda)$, with
$K_0$ and $I_0$ the modified Bessel functions~\cite{AbSt70}. Far from
the colloid, the energy density has to remain finite and integrable,
so that $h_2=h_4=0$. The integration constants $h_1$ and $h_3$ easily
follow from the boundary conditions requiring the profile and its
slope to be continuous at the point of detachment.  The minimum energy
of the functional in eq.~(\ref{functional}) is then found to be
\begin{equation}
  \label{energymin}
  E_\romfree =\pi \kappa \frac{ a}{\lambda} \left(\frac{k^3}{1-k^2}\right)
  \frac{K_0\left(ka/\lambda \right)}{K_1\left(ka/\lambda \right)} \ ,
\end{equation}
where we introduced the abbreviation $k=\sqrt{z(2-z)}$.

\begin{figure}
\includegraphics[scale=1]{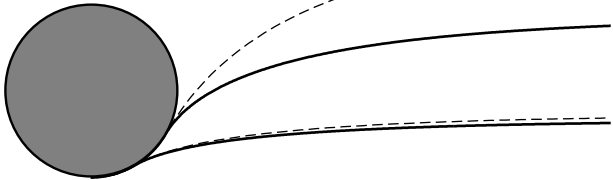}
\caption{Exact membrane profiles (solid curves) and small gradient
  approximation (dashed curves) for two prescribed detachment angles
  $\alpha = 30^\circ$ and $\alpha = 60^\circ$ for the tension $\sigma
  a^2/\kappa=0.1$.}\label{fig:profile_comparison}
\end{figure}

\begin{figure}
\includegraphics[scale=0.85]{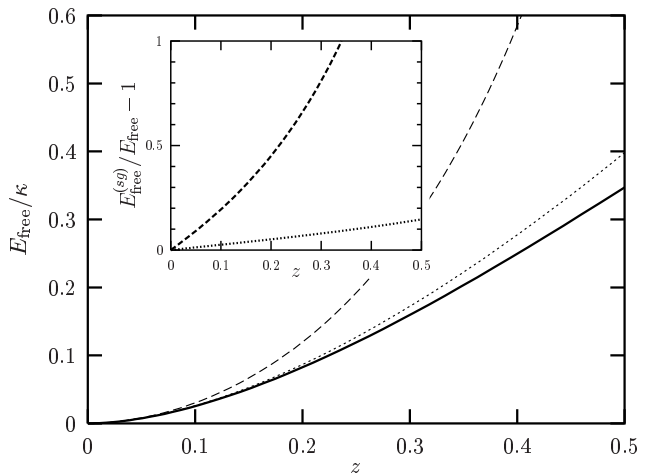}
\caption{Energy $E_\romfree$ of the free part of the membrane for the
  tension $\sigma a^2/\kappa=0.1$ as a function of the degree of
  wrapping, $z$.  Solid, dashed, and dotted curves correspond to the
  full nonlinear solution, the small gradient approximation, and its
  quadratic expansion, respectively.  The inset shows the relative
  error of the two small gradient expressions.}\label{fig:energy_comparison}
\end{figure}

The small gradient expansion and the exact membrane profiles are shown
in Fig.~\ref{fig:profile_comparison} for different values of the
detachment angle.  As expected, the quality of the approximation
declines as the degree of penetration increases (note that the
parameterization itself fails much later, namely for $z>1$).  To
further illustrate the discrepancy between the exact and the
approximate solution, we plot in Fig.~\ref{fig:energy_comparison} the
energy of the free part of the membrane.  The inset shows the energy
difference between the small gradient expansion and the nonlinear
result.  We see that the approximation is fairly reasonable for
degrees of penetration up to $z \approx 0.1$.  Beyond this value, one
has to solve the full shape equation in order to give an accurate
description of the deformation energy.

Nevertheless, the small-gradient expansion improves our understanding
of the system, since it allows us to work out the asymptotic phase
boundaries.  We consider first the transition between the free and the
partially wrapped state. For small degrees of penetration the total
energy of the system can be expanded up to quadratic order in $z$,
giving
\begin{eqnarray}
\label{smallz}
 E & = & 
 \pi a^2 \bigg\{ -2z\big(w-w_\romc\big) + 
 \\
 & & \quad\qquad \sigma z^2 \left[1-4\gamma
     -2  \, \ln\frac{\sigma  z}{w_\romc}\right] + \CALO(z^2) \bigg\} \ ,
 \nonumber
\end{eqnarray}
where $\gamma=0.5772\ldots$ is the Euler-Mascheroni constant
\cite{AbSt70}.  For sufficiently small $z$ the quadratic term is
negligible and wrapping sets in as soon as the prefactor of the linear
term becomes negative.  This confirms the numerical finding that the
initial wrapping transition is \emph{continuous} and always takes
place at $w=w_\romc$, no matter what the value of the tension is.

Even though the final state of the second transition toward full
envelopment cannot be described within a small gradient expansion, its
\emph{energy} is exactly known, since the free membrane does not
contribute to it.  The location of this second transition can
therefore also be predicted by equating the energy of the partially
wrapped state, eq.~(\ref{smallz}), with the energy of the fully
enveloped state, $E=4\pi a^2(\sigma-w)+8\pi\kappa$.  Since this yields
a transcendental equation, the relation between $\sigma$ and $w$ on
the phase boundary cannot be expressed in terms of simple functions.
However, the asymptotic behavior at zero tension, which remains quite
accurate over the whole regime of low tension, is surprisingly simple:
We find that, up to a complicated but small logarithmic correction,
the transition is given by $\sigma = w-w_\romc$ (see
Fig.~\ref{fig:pd}).

Unfortunately, the concomitant energy barrier cannot be obtained in
small gradient expansion.  Even if the transition toward envelopment
takes place at small $z$, where the approximation still works, the
barrier occurs at large penetrations, $z>1$ (see \eg\ inset of
Fig.~\ref{fig:barrier}).  This is even beyond the range of
applicability of the Monge parameterization $h=h(x,y)$.


\section{Discussion}

The combination of both numerical an analytical approaches presented
in this work has provided a complete description of the equilibrium
wrapping behavior.  The structure of the wrapping complex, the
transition lines in the wrapping diagram, and the barriers of the
discontinuous envelopment boundary can be predicted, and the low
tension regime can even be treated analytically.  One particular
consequence of the wrapping scenario is that colloid engulfment is
extremely sensitive to the particle size.  As long as the tension is
low enough ($\sigma<w/2$ will do), a large sphere is wrapped much more
easily than a small one.  This could be checked experimentally for
instance by spreading a bilayer across a hole separating two
compartments and adding a polydisperse colloidal solution to one side.
If there exists some generic attraction between the colloids and the
bilayer, a population of bilayer-coated colloids should emerge in the
other compartment with a different polydispersity distribution: The
frequency of large colloids is significantly enhanced, and too small
ones do not occur at all.

An important issue that we have not addressed here is the effect of
thermally excited membrane fluctuations. At very low tension, the
undulations might prevent the beads from adhering to the membrane.  An
initial study of finite-size effects on the Helfrich repulsion shows
that the (entropic) barrier to overcome in order to reach the surface
\emph{increases} with particle size \cite{Bickel}.  This would
provide an interesting counterbalance to the tendency to engulf
preferentially large spheres.

The level of description in this work is that of generic physical
mechanisms, but several examples mentioned in the introduction are
taken from biology, where a host of different effects often occur at
the same time.  Notwithstanding this difficulty, we believe it to be
worthwhile to analyze these examples -- in particular, viral budding
-- in terms of the mechanisms discussed here.  For instance, the
adhesion strength of viral capsids to membranes can be modified by
changing the biochemistry or simply the chemical potential of the
binding proteins which in most cases are responsible for the
attachment.  Moreover, the membrane tension of cells increases or
decreases during times of heightened endo- or exocytosis activity,
respectively, and is generally under active control of the cell
\cite{MoHo01}.  Thus, processes like viral budding and its later
counterpart of unwrapping (after infection of a different host cell
and fusion with its membrane) invariably involve a movement in the
structural phase diagram of Fig.~\ref{fig:pd} along both axes.  The
concomitant ``phase changes'' predicted that way thus offer an
understanding of the observed (un)wrapping processes in terms of the
physical forces driving them.


\acknowledgments

We would like to thank Bill Gelbart, Shelly Tzlil, and Avinoam
Ben-Shaul for many stimulating discussions on the subject.  MD also
gratefully acknowledges financial support of the German Science
Foundation under grant De775/1-1.



\end{document}